\documentstyle{mn}
\begin{document}

\input{psfig.sty}
\newcommand{\pmin}{P_{\rm min}}
\newcommand{\pturn}{P_{\rm turn}}
\newcommand{\jgr}{J_{\rm GR}}
\def\lsun{{L_{\odot}}}
\def\msun{{\rm M_{\odot}}}
\def\msol{{\rm M_{\odot}}}
\def\te{T_{\rm eff}}
\def\rsun{{R_{\odot}}}
\def\be{\begin{equation}}
\def\ee{\end{equation}}
\def\m2i{M_{2,\rm i}}
\newcommand{\simgr}{\ga}
\newcommand{\simle}{\la}
\newcommand{\lta}{\la}
\newcommand{\gta}{\ga}
\input{epsf.sty}
\def\plotone#1{\centering \leavevmode
\epsfxsize=\columnwidth \epsfbox{#1}}
\def\plottwo#1#2{\centering \leavevmode
\epsfxsize=.45\columnwidth \epsfbox{#1} \hfil
\epsfxsize=.45\columnwidth \epsfbox{#2}}
\def\plotfiddle#1#2#3#4#5#6#7{\centering \leavevmode
\vbox to#2{\rule{0pt}{#2}}
\includegraphics{#1}}

\title{Brown Dwarfs and the Cataclysmic Variable Period Minimum}
\author[U. Kolb, I. Baraffe]{U.~Kolb$^1$\thanks{Present 
address: Department of Physics, The Open University, Walton Hall,
Milton Keynes, MK7~6AA.}  
and I.~Baraffe$^{2}$\\ 
$^1$ Astronomy Group, University of Leicester,
Leicester, LE1~7RH\\ 
$^2$  Ecole Normale Sup\'{e}rieure de Lyon, C.R.A.L.\ (UMR 5574 CNRS),
F-69364 Lyon Cedex 07, France} 

\date{June 1999}

\maketitle

\begin{abstract}
Using improved, up--to--date stellar input physics tested against
observations of low--mass stars and brown dwarfs we calculate the
secular evolution of low--mass donor cataclysmic variables (CVs),
including those which form 
with a brown dwarf donor. Our models confirm the mismatch between
the calculated minimum period ($\pmin \simeq 70$~min) and the observed
short--period cut--off ($\simeq 80$~min) in the CV period histogram. 
We find that tidal and rotational corrections applied to the
one--dimensional stellar structure equations have no significant
effect on the period minimum.   
Theoretical period distributions synthesized from our model sequences 
always show an accumulation of systems at the minimum period, a
feature absent from the observed distribution. 
We suggest that non--magnetic CVs become unobservable as they are
effectively trapped in permanent quiescence before they reach $\pmin$,
and that small--number statistics may hide the period spike for
magnetic CVs.
\end{abstract}
\begin{keywords}
accretion, accretion discs --- binaries: close --- novae, cataclysmic
variables ---  stars: evolution --- stars: low--mass, brown dwarfs.
\end{keywords}

\section{Introduction}
\label{sec:intro}

Cataclysmic variables (CVs) are semi--detached binaries with a white
dwarf (WD) primary and a low--mass main--sequence companion that
transfers mass to the WD through Roche--lobe overflow
(e.g.\ Warner 1995).   
Orbital periods are known for about 400 systems. Two marked features
stand out in the orbital period distribution: a dearth of systems in
the range 2-3~hr, usually referred to as the ``period gap'', and a
sharp short--period cut--off at $\simeq 80$~min, the ``period 
minimum'' $\pmin$ (e.g.\ Ritter \& Kolb 1998). J0132--6554
at $P=77.8$~min currently marks the short end of the bulk of the
distribution, while the single system V485 Cen at $P=59$~min is the
notable exception. Six AM CVn--type  CVs with yet shorter periods are
interpreted as CVs with helium donors.   

In a semi--detached binary the Roche lobe filling star's mean density
$\rho$ determines the orbital period $P$ almost uniquely (e.g.\ King
1988), $P_h=k/\rho_\odot^{1/2}$ (where $k \simeq 8.85$ is only a weak
function of the mass ratio, $P_h$ is the period in hr, $\rho_\odot$
the mean density in solar units). The density increases 
for donors evolving along the 
hydrogen--burning main sequence towards smaller mass.
Approaching the hydrogen--burning minimum  mass the increasing
electron degeneracy induces structural changes such that further mass
loss reduces $\rho$.   
Hence during the donor's transition from a main--sequence star to a
brown dwarf (BD) the secular mean orbital period derivative changes
from negative to positive. This 
``period bounce'' has long been identified with the minimum period of
CVs (Paczy\'nski 1981, Paczy\'nski \& Sienkiewicz 1981; Rappaport, Joss
\& Webbink 1982; D'Antona \& Mazzitelli 1982). 

The actual period $\pturn$ at which a CV bounces depends on 
the ratio $\tau=t_{\rm KH}/t_M$ of the secondary's thermal time
$t_{\rm KH} \sim GM^2/RL$ and mass transfer time $t_M = M/(-\dot M_2)$,
the characteristic timescales restoring and perturbing equilibrium.
If $\tau$ is small $\pmin$ is short.
Hence $\pturn$ is sensitive both to the orbital angular momentum loss
rate (which determines the transfer rate $-\dot M_2$) and to the
interior structure of the donor. 
Paczy\'nski (1981) was the first to point out 
that $\pturn$ is close to 80 min if gravitational wave radiation
drives the mass transfer. Since then, models with different input
physics have been employed to verify a quantitative agreement between
the observed $\pmin$ and the calculated $\pturn$.   
Recent calculations (Kolb \& Ritter 1992; Howell, Rappaport \&
Politano 1997) notoriously give $\pturn$ too short by typically 10\%.

Most of these evolutionary calculations are based on input physics
which is rather
approximate for very low mass (VLM) objects, i.e.\ donors with mass
$\la \, 1 \msol$. Because of their relatively high central
densities and low central temperatures, correlation effects between
particles dominate and must be taken into account in the equation of
state (cf. Chabrier \& Baraffe 1997, and references therein).   
Below about $\te \simeq 4000$~K ($M \simle 0.6 \msol$), molecules 
become stable and dominate the atmospheric opacity, being responsible
for strong non-grey effects and significant departure of the spectral
energy distribution from blackbody emission (cf.\ Allard et al.\
1997, and references therein). 

Because of significant efforts devoted to the complex physics of 
low--mass stellar/substellar objects, the theory describing them
has improved considerably in the past few years. As a result,
the latest generation of stellar models for low--mass
stars and brown 
dwarfs is now able to reproduce observed properties of field M--dwarfs with
unprecedented accuracy (cf.\  Allard et al.\ 1996; Marley et al. 1996;
Chabrier \& Baraffe 1997; Burrows et al.\ 1997; Baraffe et al.\ 1998).

In this paper we use the Baraffe et al.\ models (1995, 1997, 1998,
henceforth 
summarized as BCAH) --- briefly reviewed in Sect.~\ref{models} --- to
calculate the secular evolution of CVs in the 
vicinity of $\pturn$ (Sect.~\ref{secev}). 
We test if tidally distorted
stellar models lead to a significant increase of $\pturn$ over the
value for spherical stars, as claimed by Nelson et al.\ 1985
(Sect.~\ref{tides}), and address the problem of the missing ``period
spike'' at $\pmin$ (Sect.~\ref{phisto}). 
This predicted accumulation of systems at $\pturn$ is caused by the
slow velocity 
in period space close to period bounce which increases the detection
probability. The spike has been a dominant feature in synthesized period
distributions obtained from theoretical CV population models
(Kolb 1993; Kolb et al.\ 1998), yet is absent in
the observed CV period distribution. 
In these earlier population models the donor star was approximated as
a (bi--)polytrope, and CVs forming with a donor mass smaller than
$0.10$~$\msun$ were not considered.  
Here we investigate if the period spike persists in
period  distributions synthesized  from full BCAH models. For the
first time, we include CVs which form already with BD donors, i.e.\
systems which did not evolve through period bounce.   
In Sect.~\ref{disc} alternative explanations for the missing period
spike are discussed. In particular, we focus on the difference between
magnetic and non--magnetic CVs.

\section{The secondary's internal structure}
\label{models}

Baraffe et al.\ (1998, and references therein) give a brief account of
the input physics used in the most recent low--mass star models we
apply here.
The main strengths of these models are in two areas: the microphysics
determining the equation of state (EOS) in the stellar interior, and
the outer boundary condition based on non-grey atmosphere models.  

The models employ the Saumon, Chabrier and Van Horn (1995) EOS which
is specifically calculated for VLM stars, BDs and giant planets.  
The EOS is an important ingredient for our analysis, since it
determines the mechanical structure of the donor stars, and thus their
{\it mass--radius relation}.  
Models based on this EOS have been successfully tested against
stars in detached eclipsing binary systems (Chabrier \& Baraffe 1995) 
and field M-dwarfs  (cf.\ Beuermann et al.\ 1998). The Saumon
et al.\ (1995) EOS has been successfully compared
with recent laser-driven shock wave experiments performed at
Livermore. These probe the complex regime of pressure dissociation
and ionization which is so characteristic for the objects we are
studying here (cf.\ Saumon et al.\ 1998). 

The second important ingredient of our VLM star and BD models concerns
the outer boundary condition, which determines the 
thermal properties of an object and thus the {\it mass--effective
temperature (\ $\te$) relation}.  As demonstrated by Chabrier \&
Baraffe (1997), evolutionary models with a grey atmosphere outer
boundary, e.g.\ the standard Eddington approximation,
overestimate $\te$ for a given mass, and yield too large a 
minimum hydrogen burning mass (MHBM). 
Because of tremendous recent progress in the field of cool atmosphere
models (see e.g.\ the review of Allard et al.\ 1997) these now provide
realistic atmosphere profiles and synthetic spectra which we use as a
more realistic outer boundary condition. 

Several observational tests confirm the success of evolutionary models
based on these improvements, e.g.\ mass--magnitude relationships,
colour--magnitude diagrams (Baraffe et al. \ 1997, 1998), mass--spectral
type relationships (Baraffe \& Chabrier 1996), and the first cool BD
GL~229B (Allard et al.\ 1996). Although some discrepancies between models
and observations remain (see Baraffe et al.\ 1998),
uncertainties attributed to the input physics now seem significantly 
reduced.   
 
We employ these one--dimensional BCAH models in the context of CVs
within the usual Roche approximation. The mass transfer rate is
calculated as a function of the difference between donor radius and
Roche radius, following Ritter (1988).   
In a separate set of calculations we apply tidal and rotational
corrections to the one--dimension stellar structure equations (Chan \&
Chau 1979) to account for the non--spherical shape of Roche
equipotential surfaces. This allows us to test the results by Nelson 
et al.\ (1985), who found that their distorted stellar models give a
value for $\pturn$ which is about $\simeq 10\%$ longer than for their 
spherical stellar models.

\section{Model calculations}

\subsection{Evolutionary sequences}
\label{secev}

Using the BCAH stellar evolution code we recalculated the secular
evolution of short--period CVs for various initial component masses.
We assume that orbital angular momentum is lost by gravitational wave
emission (e.g.\ Landau \& Lifshitz 1958) and via an isotropic
stellar wind from the WD. The wind removes the accreted mass with the
WD's specific orbital angular momentum from the binary. Hence the WD 
mass is constant throughout the evolution.  

As we focus on the period minimum we do not consider the
evolution at periods longer than 2-3~hr. In a second paper (Baraffe \&
Kolb 1999; see also Kolb \& Baraffe 1999) we discuss properties of
such sequences with more massive 
donors. They depend on the rather uncertain strength of the magnetic
braking thought to dominate the evolution of CVs above the period gap
(see e.g.\ King 1988 for a review). 

\begin{figure}
\plotone{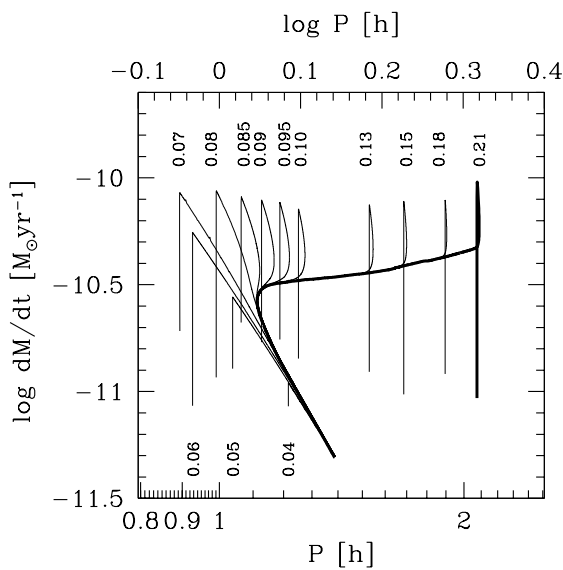}
\caption{
Mass transfer rate versus orbital period for sequences of Set A (white
dwarf mass 0.6~$\msun$). Labels indicate the initial donor mass (in
$\msun$).
\label{fig1}
}
\end{figure}

We calculated three sets of evolutionary sequences. 
In Set A the WD mass is 0.6~$\msun$, while the initial donor masses
range from 0.27~$\msun$ to 0.04~$\msun$ (see Fig.~\ref{fig1}). Set B 
is similar but for a WD mass $1.0$ $\msun$. In Set C the initial donor mass
is fixed at $0.21$~$\msun$, while the WD masses range from 1.2~$\msun$
to 0.3~$\msun$; see Tab.~\ref{tab1b}. 
At turn--on of mass transfer the secondary was either on the
ZAMS ($M_2 \ge 0.085$~$\msun$), or had an age of 2 Gyrs ($M_2 <
0.085$~$\msun$).     
All donors are hydrogen--rich and have solar metallicity ($X=0.70,
Z=0.02$). The sequences were terminated when the donor's
effective temperature $T_{\rm eff}$ was smaller than 900~K. This is
the lower limit of the temperature range covered by the present
non-grey atmosphere models (Hauschildt et al.\ 1999)
available for this study. 

\begin{figure}
\plotone{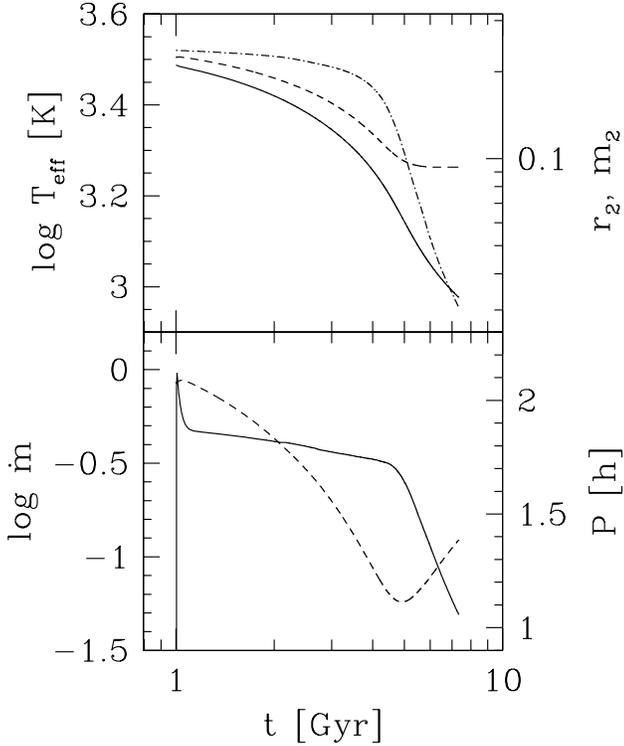}
\caption{
Time evolution of the sequence with $0.6\msun$ WD mass,
$0.21\msun$ initial donor mass sequence. The time $t$ at turn--on of mass
transfer has been set arbitrarily to 1~Gyr.
{\em Upper panel:}
Effective temperature $T_{\rm eff}$ (dash--dotted; left-hand
scale) and radius $r_2$ (dashed), mass $m_2$ (solid), both in solar units,
right--hand scale, of the donor.  
{\em Lower panel:}
Mass transfer rate $\dot m$ (in $10^{-10}$ $\msun$~yr$^{-1}$; solid)
and orbital period $P$ (dashed).
\label{fig3}
}
\end{figure}

Some quantities of interest for a typical sequence starting
immediately below the period gap are shown in Fig.~\ref{fig3} and
given in Tab.~\ref{tab2a} (see also Tab.~\ref{tab2b} for a sequence
initiating mass transfer from a BD donor).

\begin{table}
\caption{
\label{tab1b}
Model parameters of sequences in Set C (initial donor mass 0.21 $\msun$;
age of donor at birth of CV 0.61~Gyr). $M_1$ is the WD mass, $t_f$ the
age since formation of the CV when $\te=900$~K, $M_{\rm turn}$ is the
donor mass at period bounce, $\dot m_{\rm turn}$ the transfer rate
at period bounce (in $10^{-11}\msun$~yr$^{-1}$).
}
\begin{tabular}{lllll}
\hline
$M_1$($\msun$) & $t_f$ (Gyr) & $\pturn$ (hr) &
$M_{\rm turn}$ ($\msun$) & $\dot m_{\rm turn}$ \\
\hline
0.30  & 8.13  & 1.068 & 0.0634 & 1.94 \\
0.35  & 7.77  & 1.078 & 0.0633 & 2.11 \\
0.40  & 7.43  & 1.087 & 0.0631 & 2.21 \\
0.60  & 6.35  & 1.113 & 0.0628 & 2.65 \\
0.70  & 5.94  & 1.123 & 0.0627 & 2.84 \\
0.80  & 5.61  & 1.133 & 0.0624 & 3.03 \\
1.00  & 5.09  & 1.149 & 0.0622 & 3.36 \\
1.20  & 4.70  & 1.162 & 0.0620 & 3.64 \\
\hline
\end{tabular}
\end{table}

\begin{table}
\caption{Characteristic quantities along the sequence with initial
donor mass 0.21 $\msol$ and WD mass 0.6 
$\msol$. $t$ is the age in Gyr, $P_h$ the orbital period in hours,
$M$ the donor mass in $\msol$; $\te$ the effective temperature in
K; $L$ corresponds to $\log \, L/L_\odot$; $R$ is the radius in 
solar units, and $\dot M$ the transfer rate in $\msol {\rm yr^{-1}}$.
\label{tab2a}
}
\begin{tabular}{ccccccc}
\hline\noalign{\smallskip} 
$t$ (Gyrs) & $P_h$ & $M$ & $\te$ & $L$ & $R/R_\odot$ & $\log \dot M$ \\
\noalign{\smallskip}
\hline\noalign{\smallskip}
  0.00&  2.080&  0.2103& 3311.& -2.272&  0.2242&  -16.00\\
  0.25&  2.038&  0.1975& 3288.& -2.313&  0.2168&  -10.34\\
  0.50&  1.970&  0.1862& 3264.& -2.363&  0.2077&  -10.35\\
  0.75&  1.902&  0.1753& 3239.& -2.415&  0.1987&  -10.37\\
  1.00&  1.832&  0.1648& 3212.& -2.470&  0.1898&  -10.38\\
  1.25&  1.764&  0.1545& 3177.& -2.530&  0.1810&  -10.39\\
  1.50&  1.696&  0.1446& 3133.& -2.596&  0.1725&  -10.41\\
  1.75&  1.628&  0.1351& 3087.& -2.665&  0.1640&  -10.43\\
  2.00&  1.556&  0.1259& 3044.& -2.736&  0.1554&  -10.44\\
  2.25&  1.483&  0.1169& 2993.& -2.815&  0.1469&  -10.45\\
  2.50&  1.409&  0.1082& 2929.& -2.904&  0.1384&  -10.46\\
  2.75&  1.336&  0.0996& 2849.& -3.007&  0.1300&  -10.47\\
  3.00&  1.265&  0.0912& 2743.& -3.129&  0.1217&  -10.48\\
  3.25&  1.200&  0.0830& 2607.& -3.275&  0.1140&  -10.49\\
  3.50&  1.148&  0.0750& 2423.& -3.456&  0.1071&  -10.50\\
  3.75&  1.119&  0.0675& 2205.& -3.665&  0.1017&  -10.54\\
  4.00&  1.114&  0.0607& 1972.& -3.890&  0.0980&  -10.60\\
  4.25&  1.130&  0.0549& 1750.& -4.117&  0.0958&  -10.68\\
  4.50&  1.157&  0.0501& 1570.& -4.317&  0.0946&  -10.77\\
  4.75&  1.189&  0.0463& 1417.& -4.503&  0.0939&  -10.86\\
  5.00&  1.223&  0.0432& 1280.& -4.682&  0.0935&  -10.94\\
  5.25&  1.257&  0.0406& 1176.& -4.830&  0.0934&  -11.02\\
  5.50&  1.290&  0.0384& 1091.& -4.961&  0.0934&  -11.10\\
  5.75&  1.321&  0.0365& 1025.& -5.070&  0.0934&  -11.17\\
  6.00&  1.350&  0.0350&  968.& -5.168&  0.0934&  -11.23\\
  6.25&  1.377&  0.0336&  918.& -5.260&  0.0934&  -11.29\\
  6.35&  1.387&  0.0331&  900.& -5.295&  0.0935&  -11.31\\
  \hline
\end{tabular}
\end{table}

\begin{table}
\caption{Same as Table~\protect{\ref{tab2a}}, but for a BD CV sequence
with initial donor mass $M_2 = 0.07$~$\msol$ and WD mass 0.6 $\msol$.
\label{tab2b}
}
\begin{tabular}{ccccccc}
\hline\noalign{\smallskip} 
$t$ (Gyrs) & $P_h$ & $M$ & $\te$ & $L$ & $R/R_\odot$ & $\log \dot M$ \\
\noalign{\smallskip}
\hline\noalign{\smallskip}
  0.00&  0.895&  0.0700& 1796.& -4.141&  0.0887&  -16.00\\
  0.25&  1.041&  0.0571& 1626.& -4.282&  0.0919&  -10.47\\
  0.50&  1.118&  0.0505& 1475.& -4.444&  0.0927&  -10.68\\
  0.75&  1.172&  0.0461& 1334.& -4.617&  0.0929&  -10.82\\
  1.00&  1.218&  0.0428& 1220.& -4.770&  0.0931&  -10.93\\
  1.25&  1.259&  0.0402& 1129.& -4.904&  0.0932&  -11.02\\
  1.50&  1.295&  0.0380& 1056.& -5.019&  0.0933&  -11.11\\
  1.75&  1.328&  0.0362&  994.& -5.122&  0.0934&  -11.18\\
  2.00&  1.358&  0.0347&  943.& -5.214&  0.0935&  -11.24\\
  2.01&  1.359&  0.0346&  940.& -5.219&  0.0935&  -11.24\\
  \hline
\end{tabular}
\end{table}

\begin{figure}
\plotone{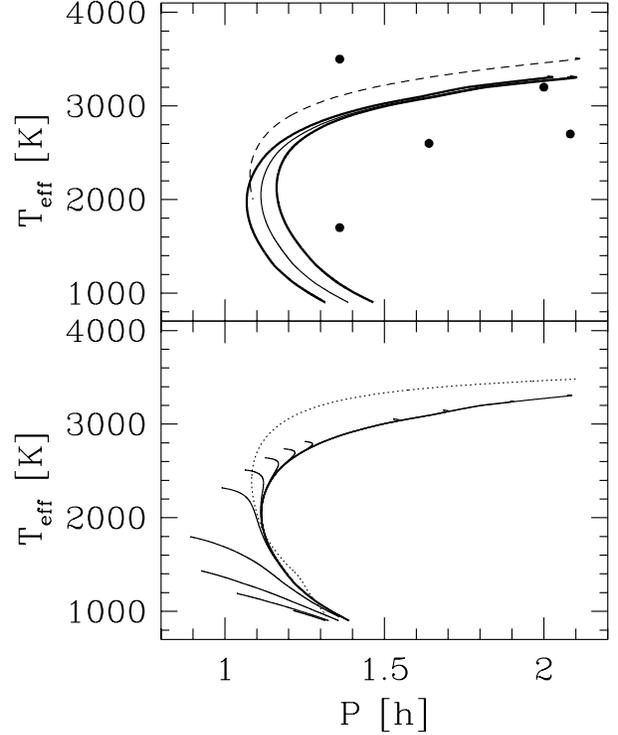}
\caption{
Effective temperature versus orbital period for various evolutionary
sequences. 
{\em Upper panel:}
Sequences with different WD masses. Solid: $1.2\msun$,
$0.6\msun$ $0.3\msun$, in order of decreasing $\pturn$. Dashed:
$0.70\msun$, but with a low--metallicity donor ($Z=0.006$). Dots show
the claimed location of AL Com (Howell et al.\ 1998), WZ Sge, TY
Psc, V592 Cas, and HU Aqr (Ciardi et al.\ 1998).  
{\em Lower panel:}
Sequences (of Set A) with different initial donor masses (solid;
see also Fig.~\protect{\ref{fig1}}). Dotted: a $0.7\msun$ WD mass sequence,
calculated with Mazzitelli's (1989) stellar evolution code.
\label{fig2}
}
\end{figure}


Fig.~\ref{fig1} confirms the well known effect that systems with
different initial donor masses rather quickly join a 
uniform evolutionary track (Stehle et al.\ 1996). Most systems of Set
A undergo period bounce at $\pturn \simeq 67$ min, which is only
slightly longer than the corresponding 
$\pturn=65$~min found with Mazzitelli's models (Kolb \& Ritter 1992;
see also the dotted curve in Fig~\ref{fig2}, lower panel). 
Note that a grey outer boundary condition like the standard Eddington
approximation yields a smaller radius for given donor mass 
$M_2 \le 0.2 \msun$, and a higher MHBM. A test calculation with fixed
mass loss rate and the Eddington approximation gives $\pturn$ shorter
by 5 min than for the corresponding sequence based on non-grey atmosphere
models.
Note also that CVs forming with fairly old and massive brown
dwarf donors (age $\ga 2$~Gyr, mass $0.05-0.07\msun$) would populate the
period regime shortwards of $\pturn$. V485 Cen might be such a system.

For Set B we have $\pturn = 69 $~min., slightly longer than for Set A
since gravitational radiation losses are higher for larger WD mass,
hence the mass transfer rate is larger. The dependence of $\pturn$ on
WD mass is most easily seen in Fig.~\ref{fig2}, upper panel.

The effective temperature of the donor as a function of period is shown in
Fig.~\ref{fig2}. The asymptotic convergence of sequences
with different initial donor mass and the insensitivity of $\dot M$
to the WD mass result in an essentially unique relation $T_{\rm
eff}(P)$ on the non--degenerate branch of the track (before period
bounce occurrs). At $\pturn$, and on the degenerate branch (after
period bounce) there is a significant spread in $T_{\rm eff}$ for 
given $P$. In the same diagram dots indicate the claimed location of AL Com
(Howell et al.\ 1998), WZ Sge, TY Psc, V592 Cas and HU Aqr (Ciardi
et al.\ 1998). For these systems the effective temperature of the donor
has been derived via spectral fittings in the near-infrared using the
same atmosphere models (Hauschildt et al.\ 1999) that serve as outer
boundary condition of our stellar models. 
Since the near--IR emission of CVs is produced by both the secondary
and the accretion disc, the estimate of the secondary's contribution
may be rather uncertain because of the lack of reliable accretion disc
models. We thus do not regard the discrepancy between model 
calculations and observational data points (cf. Fig.\ref{fig2}) as
significant.

\subsection{Sequences with non--spherical donors}
\label{tides}

We implemented the rotational correction scheme of Chan \& Chau
(1979) to investigate the effect of rotation and tidal distortion on
the donor star. Assuming solid--body rotation,
we recalculated a sequence with $0.21\msun$
initial donor mass, $1.0\msun$ WD mass and $\dot J = \dot \jgr$.  
We find that $\pturn$ is hardly affected; it is longer by only 1~min
compared to the corresponding sequence with spherical models 
(from Set B). We have performed several other experiments, with
different initial secondary masses and mass transfer rates (constant, or
as given by gravitational radiation), including the case 
considered by Nelson et al.\ (1985), i.e.\ $M_1 = 1.0 \msol$,  $M_2 =
0.4 \msol$, $\dot J = \dot \jgr$. In none of these sequences
we could reproduce the significant effect ($\sim 10 \%$)
rotational and tidal corrections had on $\pturn$ in the calculations
by Nelson et al.\ (1985) based on the same scheme of Chan and Chau
(1979).  

We do not think that the discrepancy between Nelson et al.\ (1985) and
our work results from differences in the EOS and opacities.
Indeed, although our input physics differs from those adopted  
by  Chan \& Chau (1979), we find  the same quantitative differences
quoted by these authors between the properties ($L$, $R$ and $\te$) of
spherical and rotating {\it ZAMS stars} (cf.\ their Table 1). 
We also reproduce the effects quoted in the very recent paper by Sills
\& Pinsonneault (1999). They considered a 0.7 $\msun$ ZAMS star with 
equatorial rotational velocity $145$~km~s$^{-1}$ and find a slightly
increased surface luminosity and effective temperature 
($\Delta \log L = 3.2\%$, $\Delta T_{\rm eff} = 100$~K) compared to
non--rotating models. For the same case we find $\Delta \log L =
4.5\%$, $\Delta T_{\rm eff} = 90$~K.

We thus are confident in our results and conclude that rotational and
tidal corrections as described by Chan and Chau (1979) can  be neglected
for CVs, even when the systems are close to period bounce. Note that,
although the secondary spins up in evolving toward $\pturn$, the key
quantity defined by Chan and Chau (1979), 
$$\alpha = {2 \over 3} {\omega^2R^3 \over GM_R},$$
which appears in the momentum equation
and represents the ratio of rotational to gravitational
potential gradient, remains essentially constant at
the donor star surface during the evolution (recall that the angular
velocity $\omega^2 \propto P^{-2} \propto M_2/R_2^3$). The maximum
value of $\alpha$, reached at the surface, never exceeds 0.07.  

Given the small differences between spherical and tidally distorted
models we use the less cpu--time intensive calculations with
spherical stars to derive a period distribution and analyse its
properties close to $\pmin$.

\subsection{Generating a period histogram}
\label{phisto}

\begin{figure*}
\centerline{
  \psfig{figure=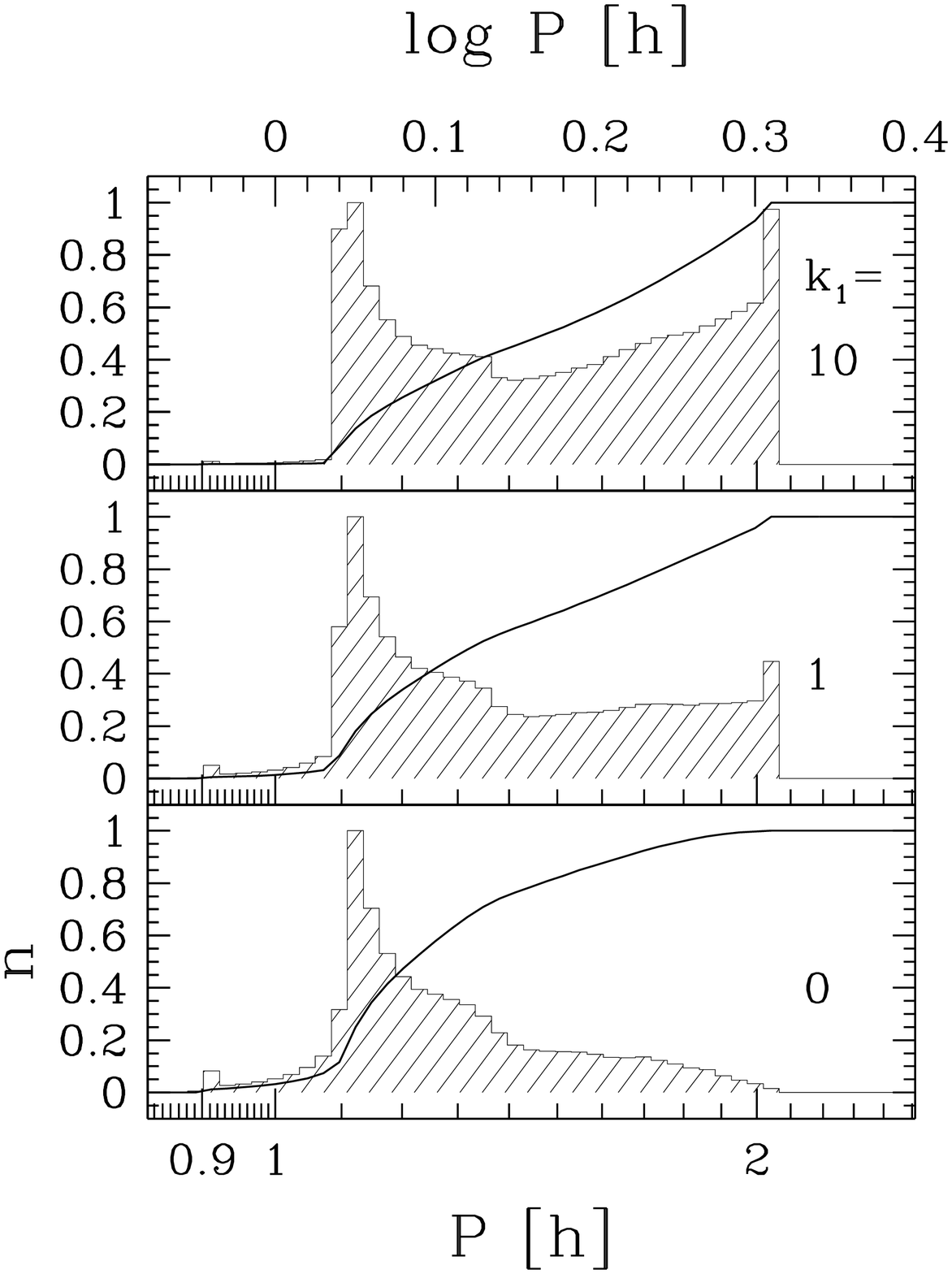,width=9cm,angle=0}
  \psfig{figure=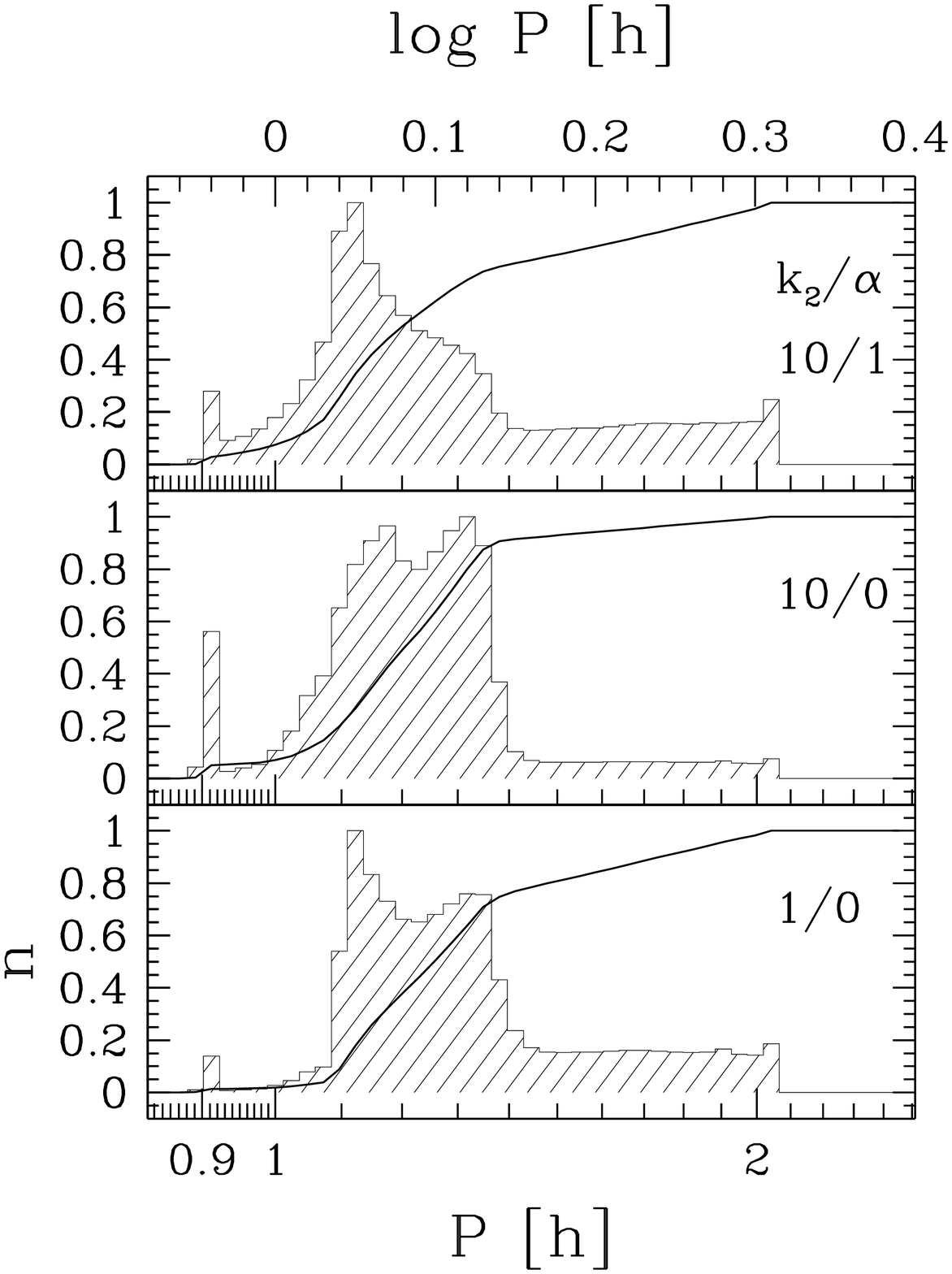,width=9cm,angle=0}
}
\caption{Orbital period distributions $n(\log P)$ for CVs with a
$0.6$~$\msun$ WD, synthesized from Set A and formation rate
(\protect{\ref{eq-b}}), arbitrarily normalised, for different 
fractions $k_1$ of systems forming above the period gap, fractions
$k_2$ of CVs forming with BD donors, and detectability parameters
$\alpha$ (see text for details).
The solid curves are the corresponding cumulative distributions
$N(>P)$.  
{\em Left:} $k_2=1$, $\alpha=1$ in all models, $k_1$ as labelled. 
{\em Right:} $k_1=1$ in all models, $k_2$ and $\alpha$ as labelled.
\label{figp1}
}
\end{figure*}

When the evolutionary sequences presented in Sect.~\ref{secev} are convolved
with an appropriate CV formation rate they give
a theoretically predicted period histogram in the vicinity of
$\pmin$ (see Kolb 1993 for a detailed description of the population
synthesis technique). 
De~Kool (1992) and Politano (1996) calculated the
formation of CVs with standard assumptions about common envelope
evolution and magnetic braking (see e.g.\ Kolb 1996 for a
review). We computed population models for all these formation rate
models; in addition, to show the differential effect of the main
parameters more clearly, we considered subpopulations for a given WD
mass with a time--independent formation rate $b(\log M_2) =$~const.\ in
certain $\log M_2$ intervals.

\smallskip

\begin{figure}
\plotone{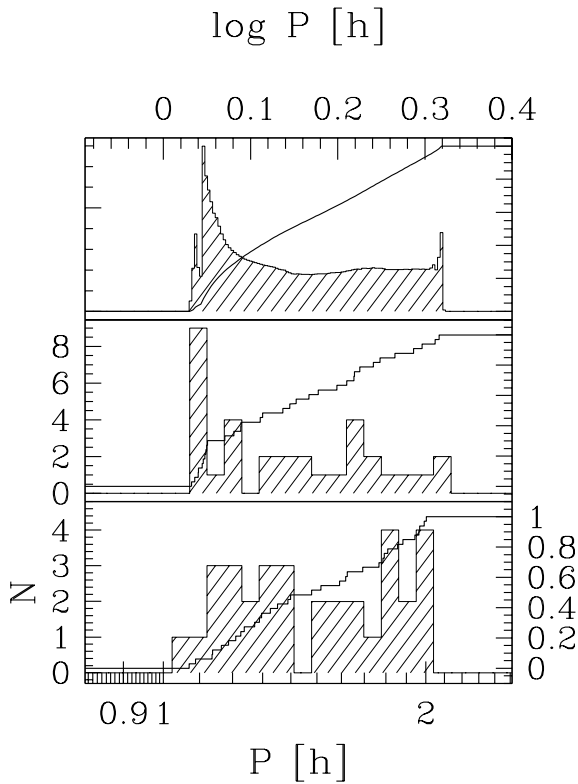}
\caption{
{\em Upper panel:} synthesized orbital period distribution P2 (see text), for
$\alpha=1.5$. {\em Middle and lower panel:}  two arbitrary samples of
33 systems drawn from the distribution in the upper panel. 
\label{figmontec}
}
\end{figure}

In Figs.~\ref{figp1}, \ref{figmontec} we show period distributions for two
different CV 
subpopulations --- P1 and P2 --- which allow one to study the main
effects that determine the structure of the histogram at $\pmin$. 
The histograms are for either a volume--limited sample,
or a sample where individual systems have been weighted by $\dot
M^\alpha$, for various values of $\alpha$. 
This mimics selection effects which affect the observed sample (cf.\
Kolb 1996).
The subpopulations formally correspond to a Galactic disc age of
6~Gyr. In a somewhat older population the edge of the volume--limited
distributions at $\simeq 1.4$~hr would appear at slightly longer $P$,
while the $\dot M^\alpha$--weighted cases would hardly change.

\paragraph*{P1.~Period histograms obtained from Set A} (Figure~\ref{figp1}).
These correspond to the period distribution of a subset of CVs with
WD mass $0.6\msun$ (assuming that this mass does not change
through the evolution). The same analysis has been performed for
sequences of Set B, with very similar results, which we therefore do
not show.  

As noted above we do not consider the evolution of CVs with
donors $>0.21\msun$, i.e.\ of CVs that form above the period
gap. In the disrupted magnetic braking model for the period gap (e.g.\
King 1988) these systems would all reappear $\la 10^8$~yr after
formation at the 
lower edge of the gap, with a donor in thermal equilibrium. 
A BCAH stellar model in thermal equilibrium with mass $\simeq
0.21\msun$ would fill its Roche lobe at $P=2.1$~hr, 
the observed lower edge of the period gap. Hence we simulate the 
contribution of systems that form above the gap to the period
distribution below the gap by increasing the formation 
rate in a narrow bin in $M_2$--space at 0.21 $\msun$ by a large factor. 

For the first time, we explicitly include systems which form with a
brown dwarf donor star. The formation rate of such BD CVs is not  
known. Survival of the common envelope phase is crucial and could be a
problem as the maximum orbital energy available to eject the 
envelope, roughly $\propto M_2$, is small.  
Simulations by Politano (1998, priv.\ comm.) show that the formation
of BD CVs is possible when the same formalism is applied as for 
more massive donor stars. For our purposes we just extrapolate the
birthrate function $b_2(\log M_2)$ down to the smallest initial donor
mass ($0.04\msun$) considered here.  

In particular, we use
\begin{equation}
b_2(\log M_2) = \left\{ \begin{array}{r@{\quad:\quad}l} 
0.368 k_1& 0.207 \leq M_2 < 0.210 \\
1 & 0.090 < M_2/\msun \leq 0.207  \\
k_2 & 0.040 \leq M_2/\msun \leq 0.090 \\
\end{array} \right. 
\label{eq-b}
\end{equation}
with various combinations of the free parameters $k_1$ and $k_2$,
describing the relative weight of systems forming above the period gap
and with a brown dwarf donor, respectively. The numerical factor in
front of $k_1$ is chosen such that for $k_1=1$ as many CVs form with
donor masses $0.21 < M_2/\msun \leq 1.0$ (``above the gap'') as with
masses $0.09 \leq M_2/\msun \leq 0.21$ (``below the gap''). The
results are not sensitive to the choice of the boundaries at
$0.207\msun$ and $0.09\msun$.
So, chiefly, $k_1$ is the ratio of systems forming above the
gap to systems forming below the gap with a non--degenerate donor,
while $k_2$ is the ratio of the formation rate of CVs with degenerate
and non--degenerate donors. 
Figure~\ref{figp1} shows period histograms for various combinations of
$k_1$, $k_2$ and $\alpha$.

\paragraph*{P2.~A period histogram obtained from Set C,} with the full
time--dependent formation rate calculated by de Kool (1992; his model
3). This 
distribution, shown in the top panel of Fig.~\ref{figmontec},
illustrates the fine structure of the period spike at $\pturn$ as a
result of the different WD masses in the sample. 
To obtain this histogram, we approximated sequences with an arbitrary
initial donor mass $M_2 <0.21$~$\msun$ by the $0.21$~$\msun$ sequence from $M_2$ 
onwards. This introduces a slight error at the onset of
mass transfer and for sequences which would form close to $\pturn$. 
An explicit comparison of results from this simplified method for
$0.6\msun$ WD mass with the distribution obtained from the full set of
sequences (set A) shows that these deviations are negligible as long
as $k_2 \la 2$. To estimate the contribution from systems that form
above the period gap the actually calculated evolutionary tracks have
been extended to donor masses $>0.21\msun$ by assuming 
a simple main--sequence mass--radius relation (such that
$P\propto M_2$) and a constant mass transfer rate ($2\times
10^{-9}$~$\msun$yr$^{-1}$).

\paragraph*{} The theoretical period distributions in Figs.~\ref{figp1},
and \ref{figmontec} show that the collective period
minimum $\pmin$ coincides with the bounce period 
$\pturn$ of individual evolutionary sequences, and is about $10$~min.\
shorter than the observed period minimum (e.g.\ Ritter \& Kolb 1998). 
This confirms tentative conclusions reached earlier by Kolb (1993) 
who considered populations constructed from simplified polytropic
stellar models with artificial outer boundary conditions, and
restricted to initial donor masses $\geq 0.09\msun$.  

The subpopulations for WD masses 0.6 $\msun$ and 1.0 $\msun$ show that
the 
period spike does not disappear or broaden significantly when CVs with
brown dwarf donor stars are included in the population. This is true
unless the majority of newly forming CVs actually are BD CVs ($k_2 \ga
5$). But if BD CVs were dominant the bulk of systems would populate
the period regime shortwards of $\pturn$, making this a fairly
unlikely possibility in view of the observed value of $\pmin$.  

In fact, in population models which do not explicitly overemphasize 
BD CVs ($k_2 \la 2$) the effect of these sequences on the overall shape
of the period spike is generally small. A fair representation of the
period spike is obtained by the simple procedure used to generate the
period distribution P2.
The double spike at $\pmin$ in the P2 histogram (Fig.~\ref{figmontec},
upper panel) is a result of the double--humped WD mass distribution
(see e.g.\ de~Kool 1992): 
high--mass C/O WD systems cluster in the longer--period spike,
low--mass He--WD systems in the shorter--period spike. This fine
structure of course disappears altogether in small samples. 

\begin{figure}
\plotone{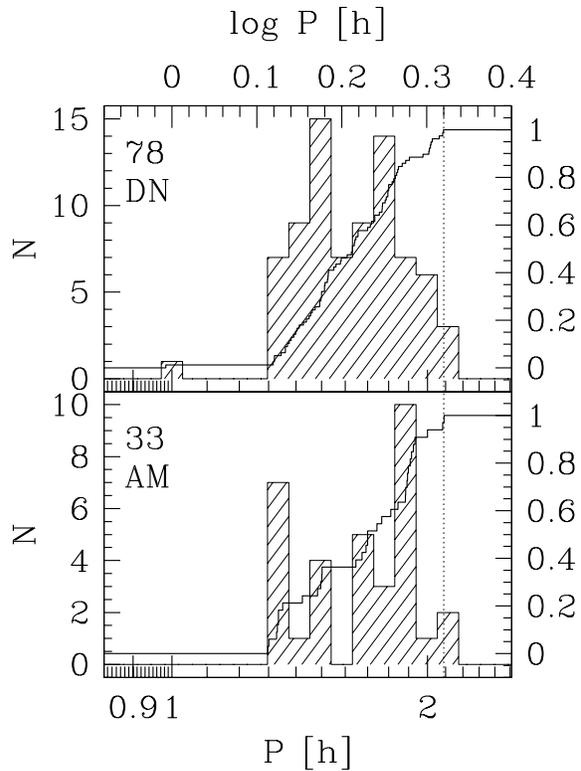}
\caption{The observed orbital period distribution of
dwarf novae (upper panel) and polars (lower panel), for 
systems with $P_h <2.1$. 
The solid curve is the normalized cumulative distribution. Data from
Ritter \& Kolb (1998). 
\label{figobs}
}
\end{figure}

The overall shape of the weighted distribution is not sensitive to
moderate $\alpha$; there is hardly any difference between histograms
calculated with $\alpha$ between $0.5$ and $1.5$. Increasing $\alpha$
further tends to decrease the amplitude of the spike. We obtain a
truly flat distribution, similar to the observed one
(Fig.~\ref{figobs}), only for $\alpha \ga 6$.

\section{Discussion}
\label{disc}

The results presented in the previous section show that 
despite the significant improvement of low--mass star and brown--dwarf
models, the period histogram for short--period CVs synthesized from
these models still appears to be in conflict with observations.   
If the evolution of short--period CVs is driven by the emission of
gravitational waves, period bounce occurs at $\simeq 70$ min for
most systems, resulting in a collective minimum period $\pmin$ which
is 10~min shorter than the observed one. The contribution of CVs
forming with BD donors is either negligible or would cause $\pmin$ to
be yet shorter. As a result of period bounce, the simulated period
histograms show an accumulation of systems near $\pmin$, the period
spike. The only way to remove the period spike effectively from the
distribution is to impose a very steep dependence of the system's
detectability $d$ on one of the evolutionary parameters, e.g.\ $\propto
\dot M^6$ or steeper. 

The catalogue of Ritter \& Kolb (1998) lists 129 CVs with periods
$<2.10$~hr. Among these are 33 polars, 78 dwarf novae (excluding 
intermediate polars), and 6 AM CVn systems. The remaining 12 systems
are either intermediate polars, novae, novalikes, or of unknown
type. Hence essentially all known short--period CVs are either dwarf
novae, i.e.\ systems with an unstable accretion disc,
or discless polars.

\subsection{The missing period spike}

Almost all dwarf novae are detected in outburst. It is tempting to
identify the accretion disc and its outburst properties as the 
cause for a strong $\dot M$--dependent selection effect.
A few short--period dwarf novae, sometimes referred to as WZ Sge
stars, have very long outburst recurrence times $t_{\rm rec}$, the
most extreme example being WZ Sge itself with $t_{\rm rec}\simeq30$~yr.
It has long been noted that low--$\dot M$ CVs might have escaped
detection if their outburst interval is significantly longer than
the period since beginning of systematic monitoring and surveying
of the sky with modern means -- a few decades.
For long $t_{\rm rec}$ the relative detection probability of
dwarf novae scales as $d \propto 1 / t_{\rm rec}$,
suggesting that $t_{\rm rec} \propto \dot M^{-6}$ or steeper. In
practice this very steep dependence would require that $t_{\rm rec}
\rightarrow \infty$ for $\dot M \la {\rm few} \times 10^{-11}
\msun$yr$^{-1}$, i.e.\ low--$\dot M$ CVs would not undergo outbursts
at all.   

Various plausible physical models which naturally account for this
property have been suggested.
Evaporation of accreted material into a hot corona could prevent the
disc from accumulating the critical surface mass density required to
launch a heating wave (Meyer--Hofmeister et al.\ 1998; see also Liu
et al.\ 1997). Systems in such a permanent quiescence would be
optically very faint but should still emit about $10\%$ of their 
accretion luminosity $L_{\rm acc} = G M_1 \dot M / R_1$ ($M_1, R_1$ is
the WD's mass and radius) in X--rays. 
Alternatively, King (1999) argues that systems approaching the
period minimum become magnetic ejectors even if the WD's magnetic
field is weak. In this case an accretion disc can no longer form,
and the systems become unobservable. WZ Sge, where the WD is suspected
to be weakly magnetic (see e.g.\ Lasota et al.\ 1999, and references 
therein), would be a marginal disc accretor (Wynn \& Leach 1999).

Although this can explain the absence of a period spike
at $\pmin$ for dwarf novae, it certainly does not apply to the
discless polars. An independent observational selection
effect operating on polars in the same period/mass transfer rate range
and with the same net result as a steep increase of $t_{\rm rec}$ for
dwarf novae seems a highly unlikely coincidence (but see
Meyer \& Meyer--Hofmeister 1999). 
In other words: polars should show a period spike, even though dwarf
novae do not. Could small--number statistics hide the period spike for
polars? 
To investigate this we performed a number of Monte Carlo experiments,
where we drew a sample of either 33 or 78 systems from an underlying
period distribution like the one shown in Fig.~\ref{figmontec} (upper
panel). As the middle and lower panels of this figure show,  
the smaller samples give a surprisingly wide variety of distributions,
with typically $20\%-25\%$ of them showing no sign of a period spike at
all. In contrast, in the larger sample the spike is prominent in more
than $90\%$ of all cases. A KS test confirms this impression: 
We wish to quantify the difference between model and observed
distribution that is due to a different morphological shape.  
Therefore we have to exclude effects arising from different values for
the lower edge $P_{\rm gap}$ of the period gap and for
$\pmin$ --- the latter effect is considered in Sect.~\ref{sec:4.2}. To
achieve this we rescale the calculated distribution such that it
matches the range of the observed distribution before performing the
KS test. Specifically, the period axis $C=\log P$ of the theoretical
distribution is rescaled according to 
\be
C \mapsto (C - O_l) \times \frac{O_u-O_l}{C_u-C_l} + O_l ,
\label{eq:rescale}
\ee
where $C_l = \log \pmin$, $C_u = \log P_{\rm gap}$ denote the
calculated period minimum and lower edge of the gap, and $O_l$,
$O_u$ the corresponding observed values. 
We applied (\ref{eq:rescale}) to the model distribution shown in
Fig.~\ref{figmontec} ($\pmin=64$~min, $P_{\rm gap}=124$~min), assuming the 
observed values $\pmin=78$~min, $P_{\rm gap}=113$~min. 
The maximum significance level for rejecting the null
hypothesis that the observed sample is drawn from this rescaled model
distribution is 0.34 for polars, 0.98 for dwarf novae. 
If we use $P_{\rm gap}=130$~min for the observed lower egde of the gap,
the rejection significance for polars rises to 0.91. This is
solely due to the so--called ``114~min spike'', see e.g.\ Ritter \&
Kolb 1992, still a significant feature in the observed distribution. 

\subsection{The mismatch between observed and calculated $\pmin$}
\label{sec:4.2}

Neither 
the small number statistics for polars 
nor 
the suggested detectability function for dwarf nova 
can make the minimum period of the   
observed distribution significantly longer than in the underlying
intrinsic distribution. 
To achieve this by the latter effect, dwarf novae have to
become unobservable long {\em before} they reach $\pturn$. 
With $\dot M$ as the most likely control parameter 
determining the detectability this is difficult to achieve, as   
$\dot M$ is almost constant on the non--degenerate branch above
$\pturn$. 
(This problem would be less severe if the main control parameter
were the donor mass, the mass ratio, or the orbital separation. 
A particularly steep dependence of the discovery probability on any of
these could be achieved if mass transfer cycles, similar to the ones
discussed by King et al.\ 1996, 1997, existed in CVs below 
the period gap, but changed their character discontinuously before 
$\pturn$.) 

To test the effect of small number statistics we drew 1000 samples of
either 33 or 78 systems from the theoretical distribution shown in
Fig.~\ref{figmontec} and registered 
the shortest period $P_s$ of each sample. The parent distribution
has a short--period cut--off at $64.4$~min. We found that $99.9\%$ of
the larger samples have $P_s<67.0$~min ($99.0\%$ have $P_s<66.6$~min,
$90.0\%$ have $P_s<65.6$~min). The median is $P_s=65$~min, the
longest $P_s$ we found is $67.03$~min. Similarly, $99.9\%$ of the smaller
samples have $P_s<71.0$~min ($99.0\%$ have $P_s<68.2$~min,
$90.0\%$ have $P_s<66.8$~min), with a median $P_s=65.4$~min, and
$71.2$~min as the longest $P_s$. 
The rare cases where $P_s$ was as long as 70~min still fall well short
of the observed value $\pmin = 78$~min.

Given this, it might be that the $\pmin$ mismatch is caused by
evolutionary effects after all --- effects not accounted for in our
models.   

The chemical composition of the secondary affects the value of 
$\pturn$, by changing the parameter of the donor at a given mass, hence
changing both $t_M$ (via the angular momentum loss time) and $t_{\rm KH}$.
Stehle et al.\ (1997) pointed out that $\pturn$ is slightly shorter for 
CVs with low--metallicity donors. They found ${\rm d}\pturn/{\rm
d}\log Z = 0.084$~hr. Our calculations confirm this ($\pturn =
67.41$~mins.\ for $Z=0.02$, while $\pturn = 64.70$~mins\ for
$Z=0.006$; both with $M_1=0.70\msun$). Although $\pturn$ increases for
higher metallicity donors the effect is much too small to account for
a mismatch of $10$~mins. 
It has been noted that a larger than expected fraction of CVs above
the gap have a nuclear--evolved secondary (Beuermann et
al.\ 1998, Baraffe \& Kolb 1999, Kolb \& Baraffe 1999). When these
secondaries become fully 
convective their helium content is $\sim 0.5$, giving a $\pturn$ which
is in fact $\simeq 7$~mins.\ {\em shorter} than for hydrogen--rich donors. 

Residual shortcomings in the EOS and the atmosphere profiles
cannot yet be ruled out as the cause for the $\pmin$ discrepancy.
A quantitative estimate of uncertainties in the mass--radius relation 
from the treatment of the EOS (see Saumon et al.\ 1995) is difficult.
The most profound observational test is against stellar parameters
determined for components in (detached) eclipsing binaries; yet to
date there are no such systems with $M_2<0.2\msun$.
The region near $\pturn$ involves effective temperatures lower than
2600~K (cf. Fig.~\ref{fig2}), below which grains form. These affect both
the atmosphere spectrum and profile. Although the atmosphere models used
for this study are grainless, calculations based on preliminary 
atmosphere models including the formation and absorption of
grains (Allard 1999; Baraffe \& Chabrier 1999) do not yield a longer  
$\pturn$. We thus do not expect an improvement of the situation with
the forthcoming generation of dusty atmosphere models.   

The $\pmin$ mismatch could also be due to uncertainties in the
calculated value of $\pturn$ inherent to the very concept of the Roche
model. Strictly valid only for point masses, its applicability to
extended donors relies on the 
fact that the stars are usually sufficiently centrally condensed. This
is not necessarily a good approximation for fully convective stars which
are essentially polytropes of index $n=3/2$. 
We note that Uryu \& Eriguchi (1999) considered stationary 
configurations of fluids describing a synchronously rotating polytropic
star in a binary with a point mass companion. For $n=3/2$ polytropes
they find a Roche radius that is typically $1-2\%$ smaller than
estimated from Eggleton's (1983) approximation for a given orbital
separation, but at the same time $\simeq 4\%$ larger than the radius
of a non--rotating polytrope with the same mass.

Alternatively, an obvious way to increase $\pmin$ is to increase the orbital 
angular momentum loss rate $\dot J$ over the value $\dot J_{\rm GR}$
set by gravitational radiation. We find $\pmin \simeq 83$~min (up from
$69$~min) for $\dot J = 4 \times \dot J_{\rm GR}$ and $M_1=1\msun$, a
much smaller increase than quoted by Paczy\'nski (1981).   
Patterson (1998) favoured a modest increase of $\dot J$ 
on grounds of space density considerations and the ratio of systems
below and above the period gap. 
Standard population models typically give a local CV space density 
of up to $10^{-4}$~pc$^{-3}$ (de~Kool 1992, Kolb 1993, Politano
1996), with $99\%$ of all CVs below the period gap, $70\%$ of these
past period bounce. Patterson argues that the observed space density
is at least a factor 20 smaller, and that there is little evidence for
the predicted large population of post--period minimum CVs. A higher
transfer rate after period bounce would remove systems from the
population as the donor can lose all its mass in the age of the
Galaxy, thus resolving both problems and the $\pmin$ mismatch.
However, postulating an as yet unknown $\dot J$ mechanism which
conspires to produce almost the same value as $\dot J_{\rm GR}$ at
the transition from non--degenerate to degenerate stars does not seem
very attractive. Rather, it seems more likely that the bulk of
short--period systems are indeed unobservable -- at least in the optical. 
Watson (1999) shows that the presence of a population of CVs with a 
space density of $10^{-4}$~pc$^{-3}$ that emits $10\%$ of its
accretion luminosity in X--rays cannot be excluded from ROSAT and ASCA
data. It should be possible to place tight limits on such a population
and its potential contribution to the Galactic Ridge emission 
from deep XMM and AXAF surveys. 

As a final note, we point out that the evolutionary effect of 
an irradiation--induced stellar wind from the donor (considered
semi--analytically by King \& van~Teeseling 1998) could help 
to resolve the $\pmin$ mismatch and the period spike problem; detailed
investigations are under way (Kolb et al.\, in preparation).

\section{Summary}

We have investigated the problem that the standard explanation for the CV
period minimum as a result of period bounce of systems 
driven by gravitational radiation predicts a period minimum that is too
short, and an accumulation of systems close to $\pturn$, the ``period
spike'', which is not observed. 

Using up--to--date stellar models by Baraffe et al.\ (1998) which
successfully reproduce observed properties of single low--mass stars
and brown dwarfs we confirm that $\pturn$ is about 10 mins.\ shorter
than the observed $\pmin$. 

We have presented synthesised period histograms for CVs below the period gap
which, for the first time, are based on evolutionary calculations
obtained with full stellar models, and include CVs which form with
brown dwarf donors. The period spike is always present.

Although there are ways to explain why the spike is not observed ---
small number statistics for polars, undetectability for dwarf
novae --- we find no satisfactory reason why $\pmin$ is longer than
$\pturn$ for both magnetic and non--magnetic CVs.  


\section*{Acknowledgments}

We are grateful to H.~Ritter, H.~Spruit, F.~Meyer,
E.~Meyer--Hof\-meister 
and K.~Beuer\-mann for valuable discussions, and 
A.~King for comments and for improving the language of 
the manuscript. 
We thank the referee, T.~Marsh, for useful comments.
IB thanks the Max--Planck--Institut f\"ur 
Astrophysik and the University of Leicester for hospitality during the
realization of part of this work. The calculations were performed on
the T3E at Centre d'Etudes Nucl\'eaires de Grenoble.

\end{document}